\documentclass[twoside]{article}
\usepackage{fleqn,espcrc2}

\usepackage{graphicx}
\usepackage[figuresright]{rotating}

\def\beq{\begin{equation}}   \def\eeq{\end{equation}}
\def\bea{\begin{eqnarray}}  \def\eea{\end{eqnarray}}
\def\nn{\nonumber}
\def\noi{\noindent}

\newcommand{\AmS}{{\protect\the\textfont2
  A\kern-.1667em\lower.5ex\hbox{M}\kern-.125emS}}

\begin{document}
\title{Discrete ambiguities in the determination of the CP angles
\thanks{Talk given by L. Oliver at the International Conference on CP
Violation Physics, September 18-22 2000, Ferrara, Italy. UAB-FT-497,
CPT-2000/PE.4073, LPT-Orsay 00-86.}}

\author{J. Charles\address{Grup de F\'{\i}sica Te\`orica \& IFAE\\
Universitat Aut\`onoma de Barcelona, E-08193 Bellaterra (Barcelona), Spain\\
\textit{on leave from:} Centre de Physique Th\'eorique (CNRS-UPR 7061)\\
Case 907, F-13288 Marseille Cedex 9, France}
\thanks{author partially supported by TMR-EC Contract No. ERBFMRX-CT980169.} 
and
        L. Oliver\address{Laboratoire de Physique Th\'eorique (CNRS-UMR 8627)\\
Universit\'e de Paris-Sud, B\^atiment 210, F-91405 Orsay, France}}

\begin{abstract}
\vspace{1pc}
We review briefly the problem of the discrete ambiguities in the determination
of the CKM angles, by classifying them into different categories. We then focus
on $\cos2\beta$ and the extraction of $\alpha$ using QCD factorization.

\end{abstract}

\maketitle

With the first measurements in $B^0_d, \bar{B}^0_d(t) \to J/\Psi K_S$ of the CP
violating parameter $\sin2\beta$ by the collaborations BaBar, BELLE \cite{1r}
and CDF \cite{2r}, the main question that arises is whether the angles  that
begin to be measured in  
experiments                                                                                                                                                         
that test CP violation match the angles that are being determined with
increasing accuracy in experiments that measure quantities that conserve CP,
i.e. the lengths of the sides of the unitarity triangle \cite{3r}. The
determination of $\gamma$ from CP-averaged $B \to K\pi,\,\pi\pi$ decay rates
\cite{4r} also enters into the latter category. This is of great importance
because it is related to the problem of the origin of CP violation, whether the
Standard Model is the correct answer or if there is some other source of it
from physics beyond the Standard Model. The question of the discrete
ambiguities (multiple solutions) in the determination of the CP angles is
directly related to the preceding issue. \par 

To begin with, let us point out that there are three types of discrete
ambiguities~: \par
 1. Ambiguities from incomplete experimental information. \par

 2. Ambiguities from the theory, with experimental information (rather)
complete, that we can call \textit{gedanken} experiments in the sense that their
solution asks for data on CP violation that can be reached only in the long
run.\par

3. Ambiguities from the theory, but with incomplete experimental
information, as is the present situation, that presumably will still continue for some
years. \par

\section{AMBIGUITIES FROM INCOMPLETE EXPERIMENTAL INFORMATION}

These are ambiguities related to the selection of the observables of some
particular process. For example, from the CP asymmetry $B^0_d$, $\bar{B}^0_d(t)
\to f$ (e.g. $f = \pi^+\pi^-$) 

\bea
\label{1e}
A_f(t)  =  {1-|\lambda_f|^2 \over 1+|\lambda_f|^2}   \cos \Delta Mt -
{2\mathrm{Im}\lambda_f \over 1+|\lambda_f|^2} \sin \Delta Mt \nn \\
\left [ \lambda_f = {q \over p} {\bar{M}(f) \over M(f)} \right
]    	 \eea

\noi one can measure 
\beq
\mathrm{Im}\left ({q \over p} {\bar{M} \over M}\right )
\mbox{ and }
\left | {q \over p} {\bar{M} \over M} \right |
\eeq
but not 
\beq\mathrm{sign}
\left [ \mathrm{Re}\left ({q \over p} {\bar{M} \over M}\right ) \right ]
\eeq
(twofold discrete ambiguity). 

	A possible solution for this kind of ambiguities is to look for processes where
the decay products decay weakly, violating parity, like 

\beq
\label{2e}
	B^0_d , \bar{B}^0_d (t)  \to \Lambda \bar{\Lambda} \to 
p\pi^{-}\bar{p}\pi^+					 \eeq

\noi Here, the more complex Lorentz structure (spin ${1 \over 2}$  of the
$\Lambda$) plus parity violation in the decay allow to reach more observables
than in the former case, and one can in principle measure \cite{5r} both
$\mathrm{Im}\left ({q \over p} {\bar{M} \over M}\right )$ and $\mathrm{Re}\left ({q \over p}
{\bar{M} \over M}\right )$. \par

	As another example of incomplete experimental information that amounts to
discrete ambiguities, let us quote the decay $B^0_d$, $\bar{B}^0_d(t) \to
J/\psi K^* \to (\ell^+\ell^-)(K\pi)_{CP=\pm}$. Looking at the time-dependent
angular distributions one can measure some correlations, but not all possible
correlations.  The observables that can be reached are, in terms of
transversity amplitudes \cite{6r}~: 

\bea 
\label{3e}
&&|A_0(t)|^2,|A_{\parallel}(t)|^2,|A_{\bot}(t)|^2, \mathrm{Re} A_{\parallel}(t)A^*_0(t),
\nn \\ 
&&\mathrm{Im}A_{\bot}(t)A^*_0(t),\mathrm{Im}A_{\bot}(t)A^*_{\parallel}(t) 
\eea

\noi Recently, Chiang and Wolfenstein \cite{7r} have pointed out that observing
the lepton polarization (i.e. muon polarization, in practice), one can reach the
missing observables 

\bea
\label{4e}
&&	\mathrm{Im} A_{\parallel}(t)A^*_0(t) , \mathrm{Re}
A_{\bot}(t)A^*_0(t) , \nn \\ 
&&\mathrm{Re} A_{\bot}(t)A^*_{\parallel}(t)		  \eea

\noi Although Ref.~\cite{7r} concerns time-integrated processes, the argument
is general since the time-dependence and the angular dependence factorize
\cite{6r}. The argument follows from the fact that the
quantities (\ref{3e}) are PT even, while the quantities (\ref{4e}) are PT odd. 

\section{AMBIGUITIES FROM THE THEORY, WITH
EXPERIMENTAL INFORMATION (RATHER) COMPLETE}

These are the ambiguities that arise from the interpretation of the observables
within a theory. One can consider  : 1) the Standard
Cabibbo-Kobayashi-Maskawa Model as the main candidate, where one has three
angles ($\alpha, \beta, \gamma$), or 2) as an example of physics beyond the
Standard Model, the Standard Model plus the possibility of a new physics (NP)
phase in the $B^0_d$-$\bar{B}^0_d$ mixing \cite{8r}, i.e.   

\beq
{q \over p}  \to e^{2i\theta_d} \left ({q \over p}\right )_{SM}						
\label{5e}
\eeq

In this case one has still three angles	($\bar{\alpha}, \bar{\beta},
\bar{\gamma}$)

\beq
	\bar{\alpha} = \alpha + \theta_d		\quad \bar{\beta} = \beta - \theta_d	\quad
\bar{\gamma} = \gamma			 
\label{6e}
\eeq

\noi But not always ($\bar{\alpha}, \bar{\beta}, \bar{\gamma}$) form a
triangle. This is a nice example of model (a devil's advocate) 
that exhibits the type of ambiguities that one can have without
contradicting the unitarity of the CKM matrix. In the $B^0_d$,$\bar{B}^0_d(t)
\to J/\psi K_S$ experiments one actually measures \cite{9r} the quantity
$\sin 2\bar{\beta}$.\par

There are two general theorems, formulated by Grossman and Quinn \cite{10r} and
recently reexamined by Kayser and London \cite{11r} \par \vskip 3 truemm

(i) If ($\sin 2\bar{\alpha}, \sin 2\bar{\beta}, \cos 2\bar{\gamma}$) are
measured [e.g. $\sin 2 \bar{\alpha}$ in $B^0_d(t) \to \pi^+\pi^-$ ignoring
penguins, $\sin 2\bar{\beta}$ in $B^0_d(t) \to J/\Psi K_S$ and $\cos
2\bar{\gamma}$ in $B^{\pm} \to DK^{\pm}$], there is a 64-fold ambiguity for
($\bar{\alpha}$, $\bar{\beta}, \bar{\gamma}$), but there is only a two-fold
ambiguity if the angles form a triangle. The remaining ambiguity can be solved
by the measurement of $\cos2 \bar{\alpha}$ or $\cos 2\bar{\beta}$, or other
quantities \cite{11r}. Note that although $\cos2\bar\gamma$ does not violate
CP, its measurement from $B^{\pm} \to DK^{\pm}$ exploits direct
CP-asymmetries.\par \vskip 3 truemm

(ii) If ($\sin 2\bar{\alpha}, \sin 2 \bar{\beta}, \sin2 \bar{\gamma}$) are
measured [e.g. $\sin2\bar{\gamma}$ in $B_s^0(t)\to\rho^0K_S$ ignoring
penguins], there is a 64-fold ambiguity for ($\bar{\alpha}, \bar{\beta},
\bar{\gamma}$), but a single solution if the angles form a triangle. \par
\vskip 3 truemm

In practice, the conditions in which the hypotheses of the above theorems are
satisfied need some word of caution: were it not for penguin contributions, the
result (i) would possibly be implemented in the long run, while (ii) is
virtually impossible because, for experimental reasons, $\sin2 \bar{\gamma}$ is
extremely difficult to measure, even assuming that penguins are negligible in
$B_s^0(t)\to\rho^0K_S$.

\section{AMBIGUITIES FROM THE THEORY, BUT WITH INCOMPLETE
EXPERIMENTAL INFORMATION}

One can interpret the data in terms of the theoretical schemes quoted
above, but assuming that only some quantities are available experimentally  with
relatively small errors. \par
	
One interesting result in this case is the following \cite{10r}~: If ($\sin
2{\alpha}, \sin 2{\beta}$) are known
(${\alpha}+{\beta}+{\gamma}=\pi$) all ambiguities are solved
knowing $\mathrm{sign}(\cos 2{\alpha})$, $\mathrm{sign}(\cos
2{\beta})$, $\mathrm{sign}(\sin {\alpha})$, $\mathrm{sign}(\sin
{\beta}$). While $\mathrm{sign}(\cos 2{\alpha})$, $\mathrm{sign}(\cos
2{\beta}$) can in principle be obtained in a model-independent way,
$\mathrm{sign}(\sin {\alpha})$, $\mathrm{sign}(\sin {\beta}$) can only
be obtained in a model-dependent way, for which some hadronic hypotheses are
necessary.

There has been
recent theoretical work on the following topics related to different types of discrete
ambiguities~: \vspace{3mm}

(i) Ambiguities linked to the measurement of $\sin 2 {\beta}$. \vspace{3mm}

(ii) Ambiguities in the exchange between CP and FSI phases. \vspace{3mm}

(iii) Looking for $\alpha$ in Dalitz plot analysis $B_d(t)\to
\pi^+\pi^-\pi^0$~\cite{SQ,12r}. \vspace{3mm} 

(iv) Ambiguities in the search of $\alpha$ in $B_d(t)\to
\pi^+\pi^-$. \vspace{3mm}

We will concentrate here on the points (i) and (iv). 

\subsection{Ambiguities linked to the measurement of $\sin 2 \bar{\beta}$}

	From the measurement of $\sin 2{\beta}$ in $B^0_d(\bar{B}^0_d)(t)
\to J/\Psi K_S$ one is left with the four-fold ambiguity
\beq
\left \{ {\beta},
{\pi \over 2}  - {\beta}, \pi + {\beta}, {3\pi \over 2}  - {\beta}
\right \}
\eeq
To lift these ambiguities one needs to measure $\mathrm{sign}(\cos
2\beta  )$ and $\mathrm{sign}(\sin \beta )$ \cite{10r}. \par

To obtain $\mathrm{sign}(\sin \beta)$, one can consider the sum of the
coefficients of $\sin \Delta Mt$ in two CP asymmetries \cite{10r,12r},
corresponding to the modes $J/\Psi K_S$ and $D^+D^-$~:

\beq
a_{\psi K_S} + a_{D^+D^-} \cong  2r_{DD} \cos \delta_{DD} \cos 2\beta \sin
\beta				 \label{7e}
\eeq

\noi where $r_{DD}$ and $\delta_{DD}$ are the modulus and phase of the ratio of
Penguin over Tree amplitudes for the mode $D^+D^-$. There is no rigorous
theoretical scheme that allows to calculate these quantities. If one assumes
factorization~\footnote{Note that the formalism of Ref.~\cite{bbns} does not
allow to establish factorization for $B\to D\bar{D}$ in the heavy quark limit.}
and knows $\mathrm{sign}(\cos 2\beta)$, then one can have an indication on
$\mathrm{sign}(\sin \beta)$. \par

There has been a significant theoretical progress concerning methods to
determine the $\mathrm{sign}(\cos 2\beta)$~: \par \vskip 3 truemm

\noi (i) Dalitz plot analyses $B^0_d(\bar{B}^0_d)(t) \to D^+D^-\pi^0$. \par

This method uses the decay chain $B^0_d(\bar{B}^0_d)(t) \to D^+D^{**-}
+D^{**+}D^- \to D^{(*)+}D^{(*)-}\pi^0$. The interferences in the Dalitz plot
allow to obtain $\sin 2\beta$ and $\cos 2\beta$~\cite{13r}. There is however an
irreductible contamination from Penguins, that behave in the same way as
current-current operators, $\Delta I = {1 \over 2}$. However, since the former
are perturbatively suppressed and one wants only the $\mathrm{sign}(\cos
2\beta)$, the method is reasonably safe. One can avoid penguin contributions by
considering the decay $B_d^0(t)\to D^+D^-K_s$, up to the price of yet poorly
known $D_s^{**}$ resonances.\par

The time dependent rate has the form, e.g. for
$\bar{B}_d^0 \to D^+D^{**-} + D^{**+}D^{-} \to D^+D^-\pi^0$

$$\left | A\left ( \bar{B}_d^0 \to D^+D^- \pi^0\right ) (t)\right |^2 = e^{-\Gamma
t {1 \over 2}} \left [ |\bar{A}_1|^2 + |\bar{A}_2|^2 \right ]$$
$$\left \{ \left ( |f^+|^2 + |f^-|^2\right ) + D \ 2 \mathrm{Re} [f^+(f^-)^*] \right .$$
$$- \cos \Delta Mt \left [ R \left ( |f^+|^2 - |f^-|^2\right ) \right .$$
$$\left . - D \sin \delta \ 2 \mathrm{Im} [f^+(f^-)^*] \right ]$$
$$- \sin \Delta Mt \left \{ D\left ( \sin (2 \beta - \delta )|f^+|^2 + \sin (2 \beta
+ \delta ) \right . \right .$$
$$\left . |f^-|^2 \right ) + \sin 2 \beta \ 2 \mathrm{Re} [f^+(f^-)^*]$$
\beq
\label{8e}
\left . \left . + R \cos 2 \beta \ 2 \mathrm{Im} [f^+(f^-)^*] \right ] \right \}\eeq

\noi where 
\beq
f(s) \sim {Y_J^0 \over s-m_R^2 + i \Pi (s)}
\eeq
 is the Breit-Wigner for the decay of $D^{**+}$ or $D^{**-}$ and the amplitudes
$\bar{A}_1 = A(\bar{B} \to D \bar{D}^{**})$, $\bar{A}_2 = A(\bar{B} \to
\bar{D}D^{**})$ correspond to $\bar{D}^{**}$ emission and to $\bar{D}$ emission
respectively. To obtain $|A(B_d^0 \to D^+D^-\pi^0)(t)|^2$ one has to change the
sign of both the $\sin \Delta Mt$, $\cos \Delta Mt$ terms. The parameters $D$,
$R$ and the strong phase $\delta$ are given by

\bea
\label{9e}
&&D = {2|\bar{A}_1|\ |\bar{A}_2| \over |\bar{A}_1|^2 + |\bar{A}_2|^2} \ R =
{|\bar{A}_1|^2 - |\bar{A}_2|^2 \over |\bar{A}_1|^2 + |\bar{A}_2|^2} \nn \\ 
&& \nn \\
&&\delta =
\mathrm{Arg}\left ( \bar{A}_1 \bar{A}_2^*\right )\eea 

\noi One sees that the $\cos 2 \beta$ term is affected by the parameter $R$.\par

 Two modellings of these decays by the Orsay group \cite{13r} (using
factorization, Isgur-Wise functions from the Bakamjian-Thomas formalism and
decay constants from lattice QCD), and the Trieste group \cite{14r} (using an
effective Lagrangian satisfying both chiral symmetry and heavy quark symmetry)
show that the contribution of $\cos 2\beta$ to the time-dependent asymmetry is
large enough to be identified. \par \vskip 3 truemm

\noi (ii) Method using $B^0_s(\bar{B}^0_s)(t) \to J/\Psi \varphi \to
(\ell^+\ell^-)(K\bar{K})$ and comparing to $B^0_d(\bar{B}^0_d)(t)
\to J/\Psi K^* \to (\ell^+\ell^-)(K\pi)_{CP}$ \cite{15r}. \par

In the $B_d$ decay chain there are angular correlations that, together with the
time-dependence, allow to measure $\cos 2 \beta \cos \delta_i$ ($i=1,2$)
\cite{5r,6r}, where $\delta_i$ are strong phases between the transversity
amplitudes. These strong phases prevent the measurement of $\mathrm{sign} (\cos 2 \beta
)$. However, in the case of the $B_s$ system, due to the sizeable lifetime
difference, there are new terms that can be measured, of the form
\beq
[e^{-\Gamma_Ht} - e^{-\Gamma_Lt}] \,\delta \varphi\, \cos \delta_i
\eeq
where, in the SU(3) limit, $\delta_i$ are the same strong phases than in the
$B_d$ case. The CP phase $\delta \varphi$, of $O(\lambda^2)$, can be measured
independently, and then $\cos \delta_i$ can be known and hence $\cos 2 \beta$.
SU(3) symmetry is a good enough hypothesis since one wants only the
$\mathrm{sign} (\cos
2 \beta )$~\cite{16r}. However, a drawback of the method is the smallness of
the CP phase $\delta \varphi$ in the Standard Model. \par \vskip 3 truemm

\noindent (iii) Measuring the muon polarization in $B^0_d(\bar{B}^0_d)(t)
\to J/\Psi K^* \to (\mu^+\mu^-)(K\pi)_{CP}$ \cite {7r}. \par
       
Since Penguins are small in this case, the new observables (\ref{4e}) would in
principle allow to measure $\cos 2\beta$. \par \vskip 3 truemm 

\noi (iv) Interference between $K_L$, $K_S$ in cascade mixing in vacuum
$B^0_d(\bar{B}^0_d)(t) \to J/\Psi K \to J/\Psi + f_K$ where $f_K$ is a common
decay mode to $K_L$ and $K_S$ \cite{17r} or using kaon regeneration in matter
\cite{18r}. \par
 
In the general case, the decay rate, dependent on the angle $\beta$, on the $B$
decay time $t_B$ and on the proper $K$ decay time $\tau_K$ writes~:

\beq
\label{10e}
\Gamma_{\beta}(t_B, \tau_K) = {1 \over 2} e^{-\Gamma_Bt_B} \Gamma(B^0 \to \Psi K^0)
\eeq 
$$e^{-N(\sigma_T + \bar{\sigma}_T)L/2} \ e^{-i\Gamma_S \tau_K}\Gamma (K_S \to
f_K)$$
$$\left \{ \left | a_{SS} + \eta a_{LS}\right |^2 \left [ 1 - \sin 2 \beta \sin
\Delta M_Bt_B\right ] \right .$$
$$+ \left | a_{SL} + \eta a_{LL} \right |^2 \left [ 1 + \sin
2 \beta \sin \Delta M_Bt_B\right ]$$
$$+ 2 \mathrm{Im} \left [ \left ( a_{SS} + \eta a_{LS}\right ) \left ( a_{SL} + \eta a_{LL}
\right )^* \right ]$$
$$\cos 2 \beta \sin \Delta M_B t_B $$
$$\left . + 2 \mathrm{Re} \left [ \left ( a_{SS} + \eta a_{LS} \right ) \left ( a_{SL} + \eta
a_{LL} \right )^* \right ] \cos \Delta M_B t_B \right \} $$

\noi where 
\beq
\eta = {A(K_L \to f_K) \over A(K_S \to f_K)}
\eeq
$\sigma_T(\bar{\sigma}_T)$ are the total cross sections of $K^0(\bar{K}^0)$,
$L$ is the length of the regenerator, and $a_{ij}$ are quantities that depend
on the regenerator parameter 
\beq
r = -{\pi N \over m_K} {\Delta f \over \Delta\lambda}
\eeq
that defines the new eigenstates in matter $K_L + rK_S$ and $K_S -
rK_L$ ($N$ is the density of scattering centers, $\Delta f = f - \bar{f}$ is
the difference of elastic forward scattering amplitudes of $K^0(\bar{K}^0)$ and
$\Delta \lambda = \Delta m_K - i \Delta \Gamma_K/2$). The off-diagonal terms
$a_{SL}$, $a_{LS}$ vanish in vacuum. In the linear approximation in $r$ one has
\bea
\label{11e}
&&m_{SS} \sim 1 \nn \\
&& m_{SL} \sim m_{LS} \sim r\left [ \exp (-i \Delta \lambda \delta
\tau ) - 1 \right ] \nn \\
&& m_{LL} \sim \exp (-i \Delta \lambda \delta \tau )
\eea
\noi where $\delta \tau = \tau_1 - \tau_2$ is the difference between the proper
times $\tau_1$ and $\tau_2$ at which the $K$ hits and leaves the regenerator. For
$\bar{B}$ decays the terms in $\sin \Delta M_Bt_B$ and $\cos \Delta M_Bt_B$ change
sign. \par 

In vacuum one has $a_{SL} = a_{LS} = L = N = \delta\tau = 0$, and a term
proportional to $\cos 2\beta$ appears with the factor $\eta$. For $f_K=\pi^+\pi^-$
the effect is very much suppressed by CP violation while  $\eta = 1$ for the
semileptonic decay $\pi^-e^+\nu_e$. However, the method suffers from the fact that
the semileptonic branching ratio is small. \par
 	
The situation is critically improved using $K_S$ regeneration in matter. In
this case, a term proportional to $\cos 2\beta$ exists even for $\eta = 0$,
that is proportional to the regeneration parameter, and one can use the
dominant decay mode $K_S \to \pi \pi$. To measure $\mathrm{sign}(\cos 2\beta)$
one would need a high luminosity $B$-factory, of the order of 300-600
fb$^{-1}$. \par \vskip 3 truemm 

\noi (vi) Measuring photon polarization in $B^0_d(\bar{B}^0_d)(t)
\to K^*\gamma^* \to (K\pi)(e^+e^-)$ decays \cite{19r}. \par

The general idea in this case is to use angular correlations between the
$(K\pi)$ and $(e^+e^-)$ decay planes to determine the helicity amplitudes in the
decay $B \to K^*\gamma^*$. The situation is similar, although in a different
region of the dilepton invariant mass, to the well-known case
$B^0_d(\bar{B}^0_d)(t)\to K^*J/\Psi \to (K\pi)(\ell^+\ell^-)$ \cite{5r,6r} where
some of the angular correlations allow to measure the product $\cos 2\beta \cdot \cos
\delta$, where $\delta$ is a strong phase difference between transversity
amplitudes. In the case examined in ref.~\cite{19r}, the lepton pair being of pure
electromagnetic origin, there is no strong phase and therefore one can in
principle measure $\cos 2\beta$.   

\subsection{The determination of the angle $\alpha$ with help of QCD factorization}

The determination of the angle $\alpha$ is no doubtly one of the major goals of
dedicated $B$-experiments. Unfortunately, the benchmark decay mode for this
angle, namely $B_d^0(t)\to\pi^+\pi^-$, suffers from small branching ratio on
the experimental side~\cite{osaka}, and of penguin contributions on the
theoretical side~\cite{gronau}.

The isospin analysis of all two-pion decay channels could in principle
disentangle penguin effects and allow the extraction of $\alpha$~\cite{gl}.
However such a study is very challenging experimentally
speaking~\cite{12r,16r}, and is plagued by high-order discrete
ambiguities~\cite{jc}.

In Ref.~\cite{jc} it was shown that estimating from theory the modulus of the
penguin-to-tree ratio with an uncertainty of $\sim$30\% at the amplitude level
may be competitive with the isospin analysis, while being much simpler. Recent
work on factorization in the heavy quark limit~\cite{bbns} suggests that such an
estimation is within the reach of our understanding. Let us recall briefly the
basic elements.

The time dependent CP-asymmetry~(\ref{1e}) of the decay is written as
\bea
A_{\pi\pi}(t)&=&a_\mathrm{dir}\cos\Delta Mt\nn\\
&-&\sqrt{1-a_\mathrm{dir}^2}\sin2\alpha_\mathrm{eff}\sin\Delta Mt
\eea
where $a_\mathrm{dir}$ is the direct CP-asymmetry while
$\sin2\alpha_\mathrm{eff}$ is an observable that reduces to
$\sin2\alpha$ in the absence of penguins. The decay amplitude is
parametrized as
\beq
A(B_d^0\to\pi^+\pi^-)=e^{i\gamma}T+e^{-i\beta}P
\eeq
where the usual phase convention for the CKM matrix is used, and where the
``tree'' amplitude $T$ receives contributions from the dominant tree diagram as
well as $u$-type and $c$-type penguin topologies, while the ``penguin'' amplitude
$P$ receives contributions from $t$-type and $c$-type penguins. We also define
the Final State Interaction (FSI) phase $\delta=\mathrm{Arg}(PT^\ast)$.

The angle $\alpha$ is related to the theoretical parameters $|P/T|$ and $\delta$
and the observables $\sin2\alpha_\mathrm{eff}$ and $a_\mathrm{dir}$ by the two
following exact equations:
\bea
&&\cos(2\alpha-2\alpha_\mathrm{eff})=\frac{1}{\sqrt{1-a_\mathrm{dir}^2}}\times\label{m1}\\
&&\times\left[1-\left(1-\sqrt{1-a_\mathrm{dir}^2}\right)\cos2\alpha_\mathrm{eff}
\left|\frac{P}{T}\right|^2\right]\nn
\eea
and
\beq\label{m2}
\tan\alpha=\frac{1-\sqrt{1-a_\mathrm{dir}^2}
\cos2\alpha_\mathrm{eff}}
{a_\mathrm{dir}/\tan\delta+\sqrt{1-a_\mathrm{dir}^2}\sin2\alpha_\mathrm{eff}}
\eeq

Once $\sin2\alpha_\mathrm{eff}$ and $a_\mathrm{dir}$ have been measured,
Eq.~(\ref{m1}) leads to four solutions for $2\alpha$ if $|P/T|$
\begin{flushleft}\begin{figure}
\includegraphics*[width=.45\textwidth]{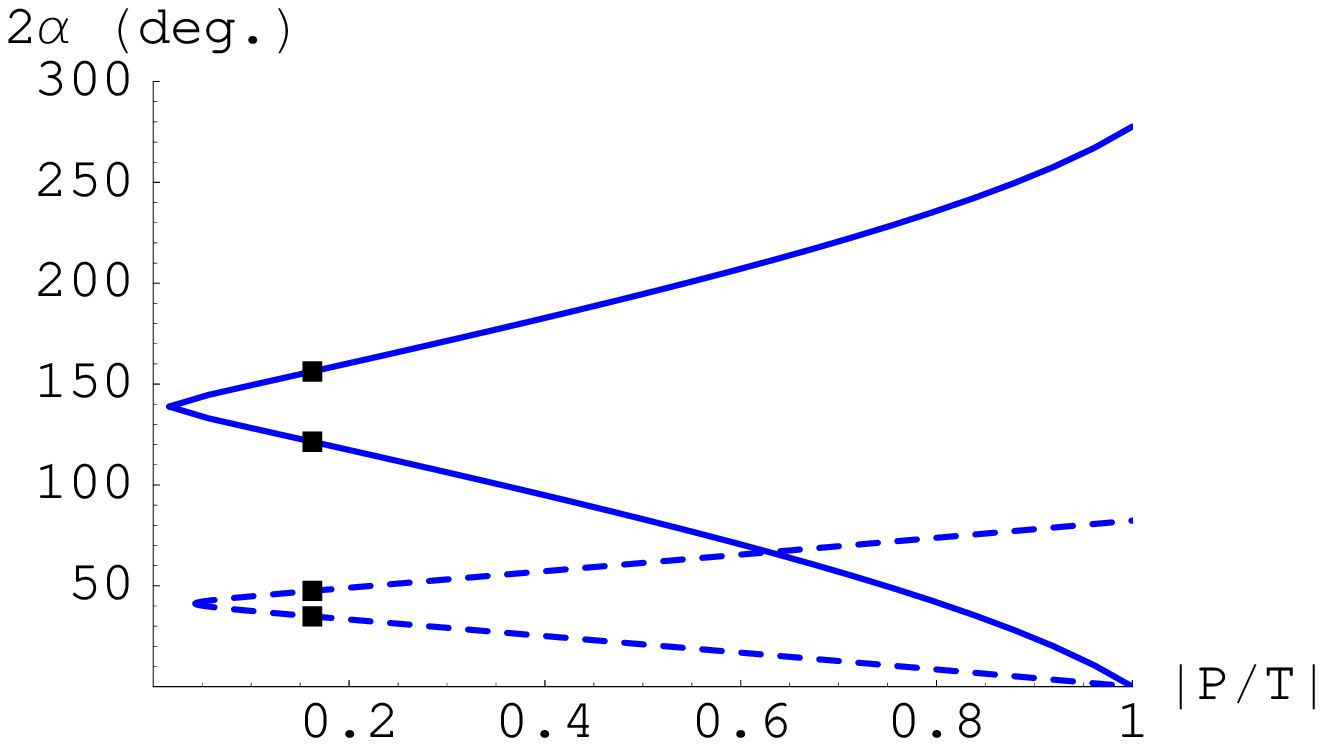}
\caption{The angle $2\alpha$ as a function of the ratio $|P/T|$ for given values of the CP
asymmetries (see text). The square dots correspond to the solutions with
$|P/T|=0.16$.\label{fig1}}
\end{figure}\end{flushleft}
is calculated
from the theory. Similarly, Eq.~(\ref{m2}) leads to two solutions for $2\alpha$
if $\delta$ is calculated from the theory. If both $|P/T|$ and $\delta$ can be
estimated with sufficient precision, then in general only one value of $2\alpha$
will satisfy both Eqs.~(\ref{m1}) and~(\ref{m2}). Furthermore, the last discrete
ambiguity $\alpha\to\pi+\alpha$ can be lifted by looking at the sign of
$\sin\delta$, thanks to the equation
\beq\label{m3}
\mathrm{sign}(\sin\alpha)\times\mathrm{sign}(\sin\delta)=\mathrm{sign}(a_\mathrm{dir})
\eeq

\begin{flushleft}\begin{figure}
\includegraphics[width=.45\textwidth]{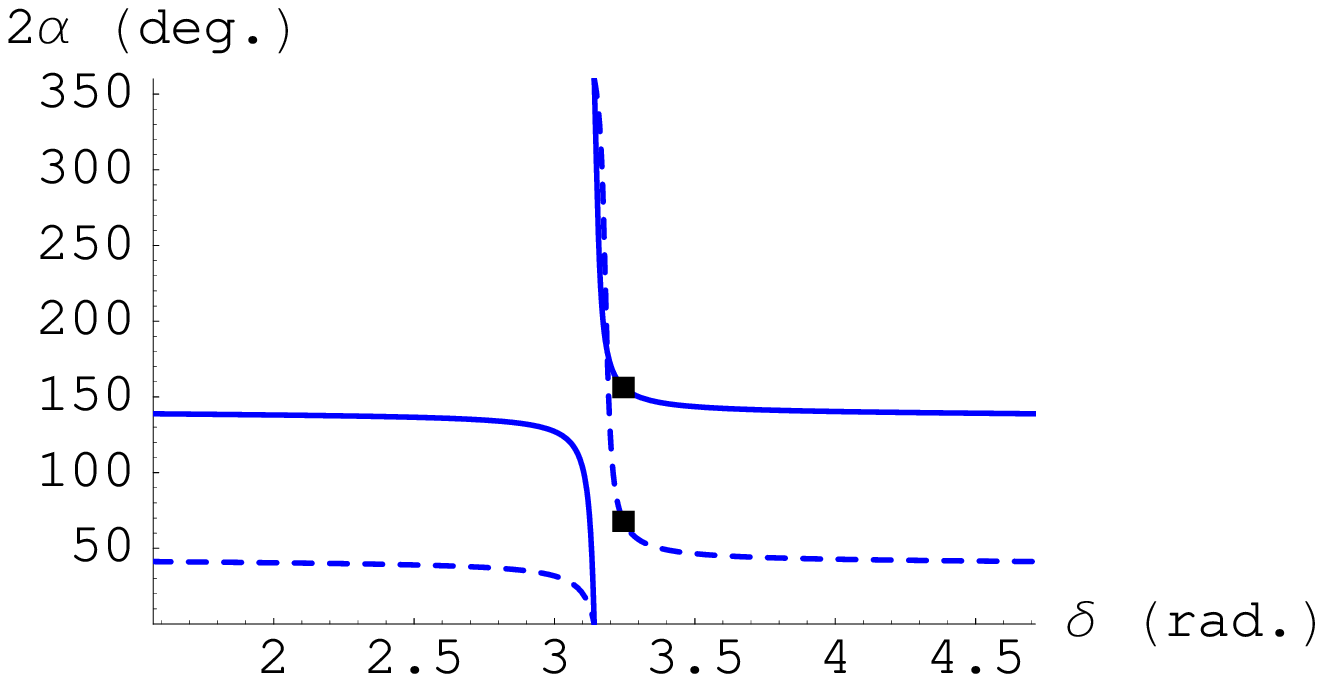}
\caption{The angle $2\alpha$ as a function of the phase $\delta$ for given values of the CP
asymmetries (see text). The square dots correspond to the solutions with
$\delta=3.25$.\label{fig2}}
\end{figure}\end{flushleft}

\begin{flushleft}\begin{figure}
\includegraphics[width=.45\textwidth]{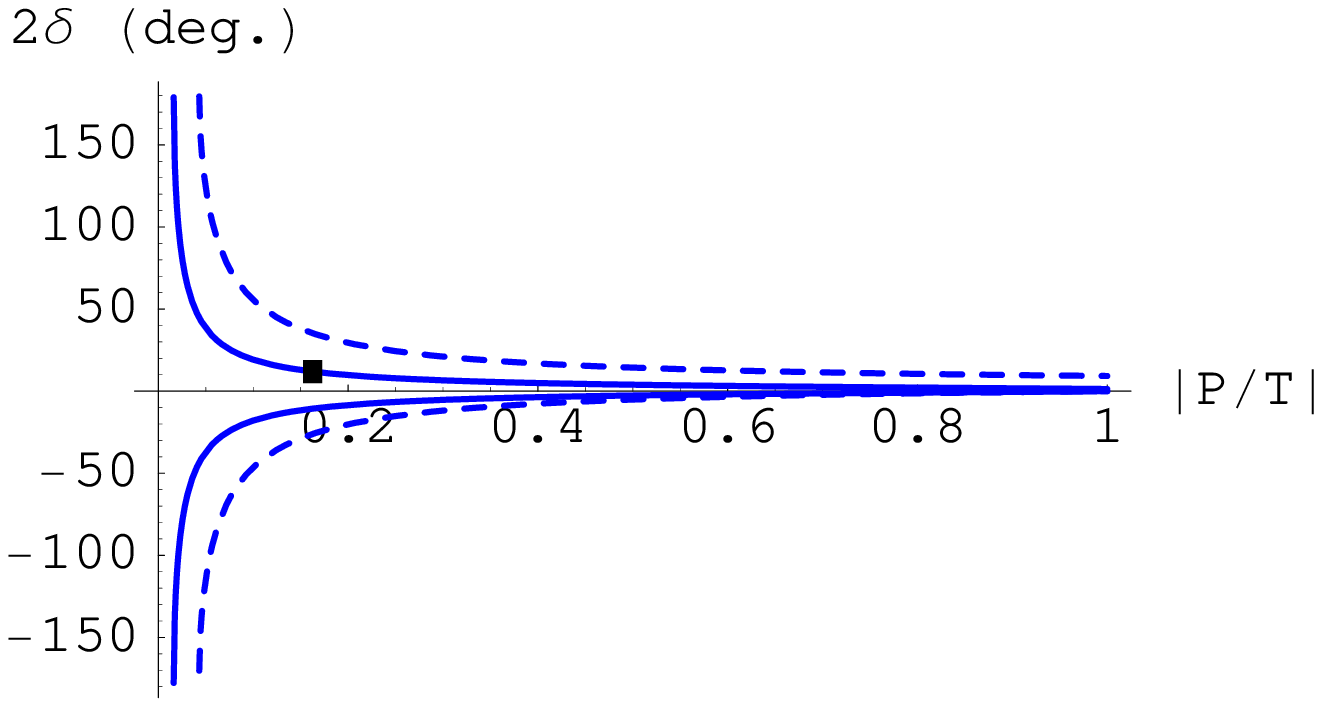}\
\caption{The phase $\delta$ as a function of the ratio $|P/T|$ for given values
of the asymmetries (see text). The square dot corresponds to
$(|P/T|=0.16,\delta=3.25)$ as given by factorization in the heavy quark limit.\label{fig3}}
\end{figure}\end{flushleft}

Recently, it has been argued that non-leptonic decays factorize in the heavy
quark limit of QCD~\cite{4r,bbns}, which should allow to compute the hadronic
matrix elements in a model-independent way, up to $1/m_b$ corrections. Assuming
for the moment that $|V_{td}/V_{ub}|$ is known, this formalism might give
explicit predictions for the parameters $|P/T|$ and $\delta$. The procedure of
how we get the angle $\alpha$ in practice is shown on
Figs.~(\ref{fig1})-(\ref{fig3}).

As an example we assume $\alpha=1.36$. We take $|V_{td}/V_{ub}|=2.43$ and we
obtain the observables $\sin2\alpha_\mathrm{eff}$ and $a_\mathrm{dir}$ from the
matrix elements calculated in Ref.~\cite{bbns}. We find
$\sin2\alpha_\mathrm{eff}=0.66$ and
$a_\mathrm{dir}=-0.03$. The latter numbers
will of course be replaced by the actual ones once the CP measurements are
available.

Let us now ``forget'' the value of the theoretical parameters ($\alpha$,
$|V_{td}/V_{ub}|$ and the QCD matrix elements) that we have used, in such a way
that we are only left with the knowledge of the two observables
$\sin2\alpha_\mathrm{eff}=0.66$ and $a_\mathrm{dir}=-0.03$: hopefully this will
be the case in a near future. The question we would like to answer is how we
could extract the value of $\alpha$ from the above observables. To this end, we
plot the contours defined by Eqs.~(\ref{m1}) and~(\ref{m2}) in the
$(|P/T|,2\alpha)$ and $(\delta,2\alpha)$ planes on Figs.~(\ref{fig1})
and~(\ref{fig2}) respectively. Note that these contours are completely
model-independent.

From $|V_{td}/V_{ub}|=2.43$ and factorization in the heavy quark limit, we find
$|P/T|=0.16$: this value is reported on the contours of Fig.~(\ref{fig1}), and
four solutions for $2\alpha$ are determined in this way. Of course we have
neglected both experimental and theoretical errors, and it remains to be
investigated with which precision $2\alpha$ can be extracted from this
approach. The uncertainty on the CKM angle depends on the experimental errors
and the theoretical error on $|P/T|$.

Similarly, factorization predicts $\delta=3.25$, the value of which is reported
on Fig.~(\ref{fig2}): this allows to select two possible values for $2\alpha$.
One sees that only one value for the CKM angle is compatible with both the
calculated $\delta$ and $|P/T|$ (the one corresponding tp the upper dot in both
figures, with $2\alpha=156^\circ$). The remaining $\alpha\to\alpha+\pi$
ambiguity is lifted as indicated above by using Eq.~(\ref{m3}) and
$\sin\delta<0$ and $a_\mathrm{dir}<0$: we are left with a single value of the
CP angle, $\alpha=1.36$, i.e. the value that we had as input for the
calculation of the observables.

The fact that redundant information contained in $|P/T|$ and $\delta$ eliminates
spurious solutions can be viewed in a different way: eliminating $\alpha$
between Eqs.~(\ref{m1}) and~(\ref{m2}), we obtain $2\delta$ as a function of
$|P/T|$, the contour of which is drawn on Fig.~(\ref{fig3}). Any estimation of
$(|P/T|,2\delta)$ should fall precisely on one of the branches: provided the
various sources of error are controlled, this plot may constitute a good test of
the theory. In case of a disagreement, this may be an indication of either the
failure of the hadronic calculation, or New Physics.

Note that the knowledge of the ratio $|V_{td}/V_{ub}|$ is likely to be affected
by still large uncertainties at the time where the CP experiments give first
data. Actually, once the Standard Model is assumed, it is possible to express
both $\alpha$ and $|V_{td}/V_{ub}|$ in terms of the Wolfenstein parameters
$\rho$ and $\eta$, and solve directly the whole problem as contours in the
$(\rho,\eta)$ plane~\cite{jc}.

This shows that with the help of the theory, the angle $\alpha$ can be
determined without any discrete ambiguity in the general case. Of course,
beside experimental uncertainties that will give some ``width'' to the contours
of Figs.~(\ref{fig1})-(\ref{fig3}), we have to handle the theoretical errors
associated with the calculation of $|P/T|$ and $\delta$, which both receive
perturbative and non-perturbative $1/m_b$ corrections: the former are in
principle calculable order by order in perturbation theory, while the latter
may be much more difficult to control~\cite{bbns}.

\subsection{Conclusion}
We have briefly reviewed part of the great theoretical progress concerning the
determination of $\cos2\beta$, and re-examined the extraction of the angle
$\alpha$ in light of the recent QCD-based work on non-leptonic $B$-decays into
two light mesons.

\subsection*{Acknowledgement}
The authors acknowledge partial support from TMR-EC Contract No. ERBFMRX-CT980169.


\begin{thebibliography}{9}
\bibitem{1r} Talks presented by G. Sciolla (BaBar Collaboration) and Y. Sakai
(BELLE Collaboration), these proceedings. 

\bibitem{2r} T. Affolder et al. (CDF Collaboration), Phys. Rev. D61 (2000)
072005.   

\bibitem{3r} Talk presented by F. Parodi, these proceedings. 

\bibitem{4r} Talk presented by M. Neubert, these proceedings; M. Beneke, G.
Buchalla, M. Neubert and C. T. Sachrajda, contribution to the conference ICHEP
2000, July 27-August 2, Osaka, Japan.

\bibitem{5r} J. Charles, A. Le Yaouanc, L. Oliver, O. P\`ene and J.-C. Raynal,
Phys. Rev. D58 (1998) 114021.  

\bibitem{6r} I. Dunietz, H. Quinn, A. Snyder and W. Toki, Phys. Rev. D43 (1991)
2193; A. S. Dighe, I. Dunietz, H. J. Lipkin and J. L. Rosner, Phys. Lett. B369
(1996) 144.  

\bibitem{7r} Cheng-Wei Chiang and Lincoln Wolfenstein, Phys. Rev. D61 (2000)
074031. 

\bibitem{8r} Y. Grossman, Y. Nir and M. P. Worah, Phys. Lett. B 07 (1997) 307.
 

\bibitem{9r} G. C. Branco, L. Lavoura and J. P. Silva, CP Violation, Oxford
University Press, Oxford, 1999. 

\bibitem{10r} Y. Grossman and H. R. Quinn, Phys. Rev. D56 (1997) 7259. 

\bibitem{SQ} A. E. Snyder and H. R. Quinn, Phys. Rev. D48 (1993) 2139.

\bibitem{11r} B. Kayser and D. London, Phys. Rev. D61 (2000) 116012.

\bibitem{12r} The BaBar Physics Book, SLAC-R-504 (1998). 

\bibitem{bbns} M. Beneke, G. Buchalla, M. Neubert and C. T. Sachrajda, Phys.
Rev. Lett. 83 (1999) 1914; hep-ph/0006124.

\bibitem{13r} J. Charles, A. Le Yaouanc, L. Oliver, O. P\`ene and J.-C. Raynal,
Phys. Lett. B425 (1998) 375 and erratum-ibid B433 (1998) 441.

\newpage

\bibitem{14r} P. Colangelo, F. De Fazio, G. Nardulli, N. Paver and Riazuddin,
Phys. Rev. D60 (1999) 033002.

\bibitem{15r} A. S. Dighe, I. Dunietz and R. Fleischer, Eur. Phys. J. C6 (1999)
647.  

\bibitem{16r} $B$ Decays at the LHC, CERN-TH/2000-101, hep-ph/0003238. 

\bibitem{17r} B. Kayser, in Proceedings of the 32nd Rencontres de Moriond:
Electroweak Interactions and Unified Theories, Les Arcs, France, edited by J.
Tr\^an Thanh V\^an, Ed. Fronti\`eres (1997), p. 389; Ya. I. Azimov, JETP Lett.
50 (1989) 447, Phys. Rev. D42 (1990) 3705. 

\bibitem{18r} H. R. Quinn, T. Schietinger, J. P. Silva and A. E. Snyder,
hep-ph/0008021 (2000). 

\bibitem{19r} Y. Grossman and D. Pirjol, J. High Energy Phys.  006 (2000) 029.

\bibitem{osaka} Results from the CLEO collaboration, presented at the conference
ICHEP 2000, July 27-August 2, Osaka, Japan; from the BaBar collaboration,
\textit{ibid.}; from the BELLE collaboration, \textit{ibid.}

\bibitem{gronau} M. Gronau, Phys. Rev. 63 (1989) 1451.

\bibitem{gl} M. Gronau and D. London, Phys. Rev. Lett. 65 (1990) 3381.

\bibitem{jc} J. Charles, Phys. Rev, D59 (1999) 054007.

\end{thebibliography}
\end{document}